# Strong coupling of in-plane plasmon modes and their control

Sachin Kasture[1], Prasanta Mandal[1,2], S. Dutta Gupta[3], Achanta Venu Gopal[1*]

[1]DCMPMS, Tata Institute of Fundamental Research, Homi Bhabha Road, Mumbai 400005, India

[2]Present Address: Department of Physics, IIT, Kanpur 208016, India

[3]School of Physics, University of Hyderabad, Hyderabad 500046, India

**Abstract :** We show anti-crossings due to strong in-plane coupling of plasmon modes in dielectric-metal-dielectric structure with top 2D dielectric pattern. Experimentally measured anti-crossing widths are compared with those calculated by coupled mode theory. It is shown that the coupling strength of the plasmon modes can be controlled by the orientation of the sample.

Strong coupling of polariton modes results in anti-crossing or avoided crossings in the dispersion plots and are interesting for entanglement generation among others. Controlling the strength of coupling and thus the anti-crossing width or split gap between the two modes is of interest. For example, anti-crossings due to exciton-photon coupling in cavities and exciton-exciton coupling in coupled quantum dots are reported [1-3]. In the context of plasmons, coupled plasmons are studied where the coupling is between localized (particle) plasmons and propagating plasmons and coupling induced changes in dispersion in planar and corrugated structures [4-7]. In various sub-wavelength structures, these coupled localized and propagating plasmons are studied for basic physics as well as for different applications [8-13]. Also, in metal-dielectric-metal or dielectric-metal-dielectric structures when the middle layer is thin enough, coupling between plasmons at the top and bottom interfaces results in splitting of the modes to symmetric (short range) and anti-symmetric (long range) plasmon modes [14]. The general case of spp excitation in 1-d system when the grating vector is not contained in the plane of incidence has also been considered [15]. Such excitation geometry would allow the coupling of in-plane plasmons in case of two dimensional gratings. In this paper we report on strong coupling of in-plane propagating plasmons at the unpatterned dielectric-metal interface originating due to the 2D planar structure on the top in the non-conical excitation geometry. In addition, we show that the coupling strength and thus the anti-crossing gap (split) width can be controlled.

The structure of the paper is as follows, we will first briefly describe the sample and the measurement geometry. Later, we show the experimental results on dispersion measurements which demonstrate anti-crossing of plasmon modes and control of the splitting. We then quantitatively explain the splitting due to the coupling of plasmon modes followed by summary.

Samples studied have dielectric-metal-dielectric layer structure with 2D air holes in dielectric pattern on top. On quartz substrates, Gold was deposited by sputtering followed by spin coating of Shipley's S1805 photoresist on top. By interference lithography shallow 2D gratings of air holes in resist with depth of air holes of about 60nm were prepared. There is an unpatterned resist of 400nm thickness on the Gold layer constituting the top dielectric layer above the metal film. AFM and SEM images showed 2D air hole pattern is of rectangular lattice with periods 610nm ($a_1$) and 625nm ($a_2$) in the two orthogonal directions. The fill factors (air to dielectric ratio) are 0.65 and 0.7 in x and y directions, respectively. When the azimuthal angle is, $\varphi=0°$, for TM polarized incident light launched in the plane perpendicular to the 2D pattern, the electric field component has finite component along $a_1$ which excites the SPPs. Similarly, for $\varphi=90°$ the finite electric field component along $a_2$ excites the SPPs. Angle $\theta$ is the launch angle in the plane perpendicular to the XY-plane. Figure 1 shows the schematic of the sample and the measurement geometry.

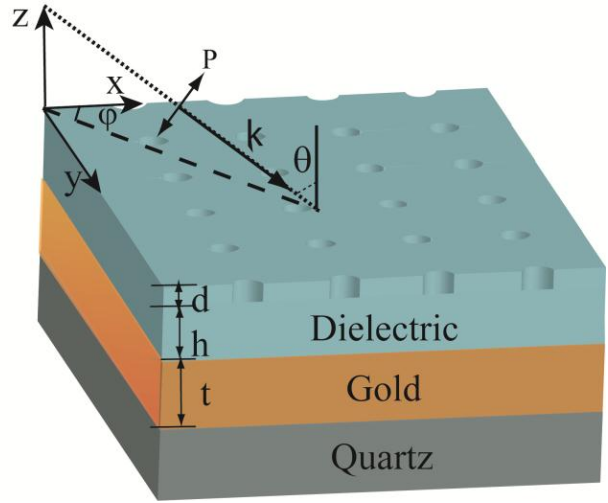

Fig.1 Shown is the schematic of the sample structure with dielectric-metal-dielectric layers having a 2-D air hole pattern in the top dielectric. Top dielectric is S1805 photoresist. Launch angle $\theta$ with respect to k and the azimuthal (in-plane) angle $\varphi$ are shown. Various parameters are the air hole depth (d), height of top dielectric (h) and the thickness of the gold layer (t) are also marked.

Angle resolved transmission measurements are performed with the launch angle varied from -25° to +25° range with 0.3° step. A 100W halogen lamp is used as white light source (400-1000nm wavelength range) and the output is collimated using a combination of apertures and lenses to have <0.3° divergence. Transmission spectra are recorded using a fiber spectrometer with 0.1nm spectral resolution.

Figure 2 shows the measured dispersion for different angles $\varphi$. We have recently reported possibility to excite plasmon modes that are independent of the launch angle ($\theta$) for $\varphi=0°$ and $\varphi=90°$ for TM polarized incident light [16]. In Fig. 2, the contour plots of the measured SPP dispersion are shown in grey scale plot thus the white regions show dips in the transmission spectra corresponding to plasmon excitation. Dashed lines are the fits. While for $\varphi \sim 0°$ and $\varphi \sim 90°$ orientations we see the flat dispersion modes, for all azimuthal angles we see the clear anti-crossing signatures which are marked by circles on the figures. In order to quantitatively explain the results and to show that $\varphi$ can be used to control the strength of the splitting, we extended the coupled mode theory to our sample and measurement geometry.

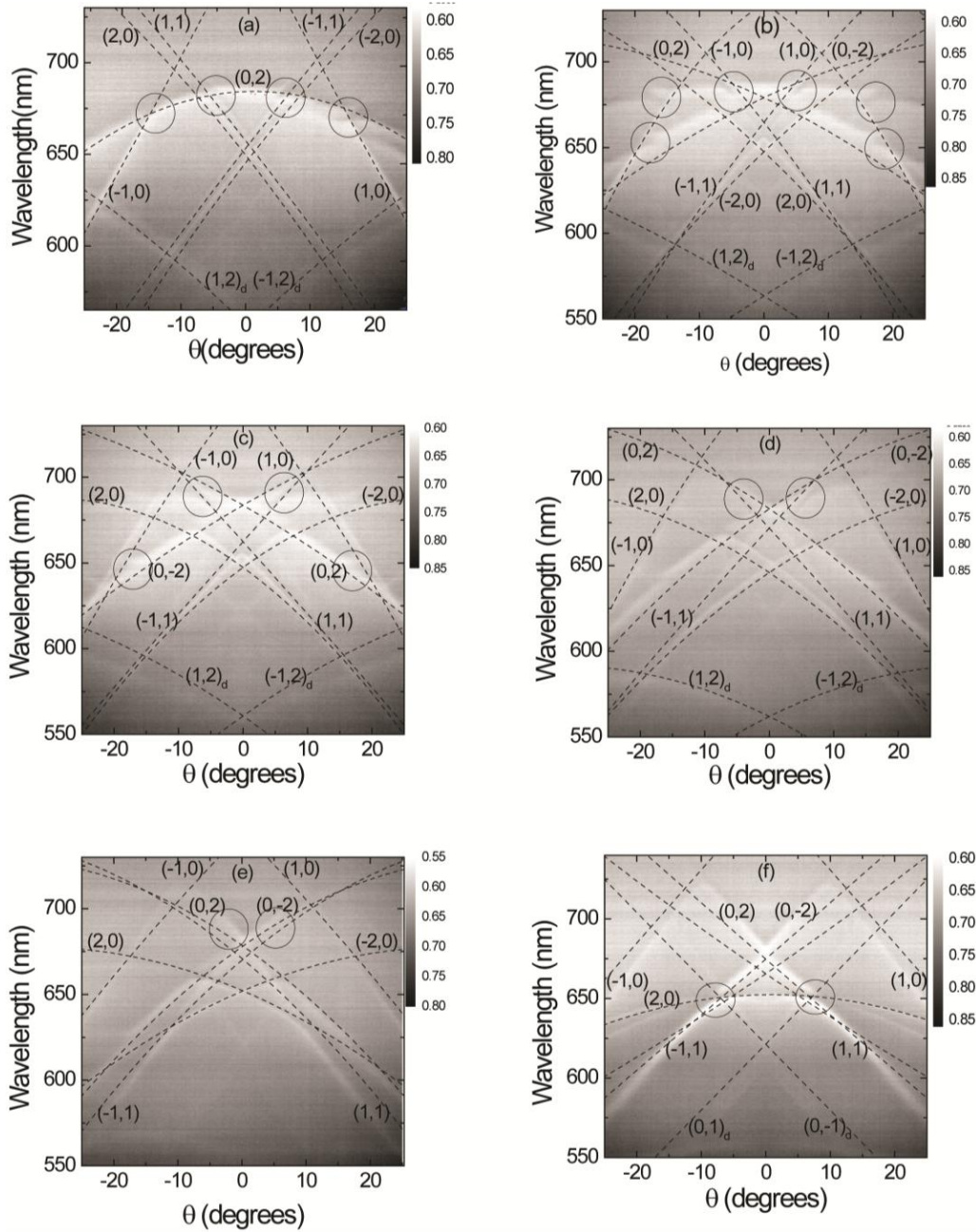

Figure 2: Measured surface plasmon dispersion in dielectric-metal-dielectric layer structure with 2-D dielectric pattern on top for different azimuthal angles is shown as grey scale plots. (a) $\varphi = 0º$, (b) $\varphi = 15º$ (c) $\varphi = 30º$, (d) $\varphi = 45º$, (e) $\varphi = 60º$ and (f) $\varphi = 90º$. Whiter regions are the plasmon absorption related transmission dips in the spectra. Dashed lines are fits (see text). Different modes are labeled. Anti-crossing sites have been marked by circles.

Coupled mode theory is well known for the orthogonal set of waveguide modes, gratings and other systems [17,18]. Briefly, when one of the modes locally modifies the dielectric constant then this perturbation induces coupling between different orthogonal modes. In such a picture for the slab waveguide geometry (infinite in x- and y-directions and finite in z-direction) for modes propagating along x-direction, the in-plane and out-of-plane coupling constants are well known. For 1-D metal grating structure, such a model has been applied recently [19]. We apply the coupled mode theory to the three layer structure with 2D pattern on top by the following procedure. We first calculate the field expressions for the plasmon (surface) modes possible in a 3-level structure and get the expressions for coupling constants for non-collinearly coupled modes. In the 2$^{nd}$ step, we introduce effect of pattern by introducing the relevant k components so that we take care of the effect of the pattern in selectively exciting the modes based on momentum matching.

In the first step, we consider the 3-level structure and the relevant expressions for the TM polarization components ($E^x$, $H^y$, and $E^z$) and calculate the coupling constants. For a three layer system, a thin metal film sandwiched between dielectric materials, solutions to Maxwell's equations with appropriate boundary conditions at the interfaces for TM polarized wave are well known [20]. In the three regions, (in the top dielectric, in the metal and in the bottom dielectric), expressions for the three relevant field components ($H^y$, $E^z$ and $E^x$) are in 3 unknown constants. We simplify these equations and write them in terms of a single constant which is solved for by using the power normalization relation,

$$\frac{1}{4} \iint (E_\alpha^{t*} \times H_\beta^t + E_\beta^t \times H_\alpha^{t*}) dydz = \pm \delta_{\alpha\beta} \cos\varphi \qquad (1)$$

where the integration extends into all the three regions along z-axis and $\varphi$ is the azimuthal angle with respect to the x-axis.

In this case, when $E(r) = \sum_\beta A_\beta(x) E_\beta(y,z) e^{ik_\beta x}$ is the total field summed over all modes, where $A_\beta$ governs the evolution of the field, and when two of the TM modes are coupled, the in-plane and normal to plane coupling constants are given similar to the overlap integrals [21].

In the 2$^{nd}$ step, we extend the model to the present structure of a three layer system with a 2-D shallow grating on top by considering a periodic perturbative correction to the dielectric constant in the x-y plane. The periodic function can be Fourier expanded as,

$$\Delta\varepsilon(x, y) = \sum_{mn} \Delta\varepsilon_{mn} e^{-imK_x x} e^{-inK_y y} \qquad (2)$$

where $K_x = 2\pi/a_x$ and $K_y = 2\pi/a_y$ with $a_x$ and $a_y$ being the periodicities in the x and y directions, respectively.

On substituting this in the field propagation equation it reduces to,

$$\frac{dA_\alpha(x)}{dx} = -i \sum_{mn} K_{\alpha\beta}^{mn} A_\beta(x) e^{-i(k_\beta + mK_x - k_\alpha)x} \qquad (3)$$

where the transverse and the in-plane coupling constants for two interacting modes, α and β, are given by,

$$K_{TM\alpha\beta}^{mn,t} = \frac{\omega\varepsilon_0}{4} \iint E_\alpha^{z*} \Delta\varepsilon_{mn} e^{-inK_y y} E_\beta^z dydz$$

$$K_{\alpha\beta}^{mn,xy} = \frac{\omega\varepsilon_0}{4} \left\{ \iint (E_\alpha^{x*} \Delta\varepsilon_{mn} e^{-inK_y y} E_\beta^x) \cos(\varphi_\alpha - \varphi_\beta) dydz \right\} \quad (4)$$

where $\varphi_\alpha$ and $\varphi_\beta$ are the angles made by the propagation vector of each of the waves with the x-axis and $E^x$ is the electric field component along the propagation direction of each mode. The total coupling constant is the sum of the transverse and in-plane coupling constants.

In the 2$^{nd}$ step, we calculate the specific k values allowed by the sample and excitation geometry to find the coupling between coplanar SPP modes. For the three layer system with appropriate boundary conditions when we invoke continuity of the field and its derivative at each interface, we get the SPP dispersion relation given by [20],

$$\tanh(k_1 t) = -\frac{\varepsilon_1 k_1 (\varepsilon_2 k_3 + \varepsilon_3 k_2)}{(\varepsilon_2 \varepsilon_3 k_1^2 + \varepsilon_1^2 k_2 k_3)} \quad (5)$$

where $k_j^2 = k^2 - k_0^2 \varepsilon_j$, j =1,2,3. In this k values are obtained from Equation (6) below which gives the possible momenta allowed by the top 2D dielectric pattern.

$$k_{in-plane} = \sqrt{(k_0 \sin\theta\cos\varphi \pm \frac{m2\pi}{a_x})^2 + (k_0 \sin\theta\sin\varphi \pm \frac{n2\pi}{a_y})^2} \quad (6)$$

The first term in the square root is the $k_x$ component and the second term is the $k_y$ component. For given structure (that is, $a_x$ and $a_y$) and the measurement geometry (θ,φ), we find the $k_{in-plane}$ values satisfying both Eq.5 and Eq.6 simultaneously. We calculate the coupling constant between two such resonant modes A1(x) and A2(x) given by,

$$A1(x) = A_\alpha(x) e^{i(k_\beta - k_\alpha + mK_x)x}$$

$$A2(x) = A_\beta(x) e^{-i(k_\beta - k_\alpha + mK_x)x} \quad (7)$$

We solve simultaneously the coupled differential equations (given by Equation (7)) for these two modes in order to get the splitting due to the coupling. For perfect phase matching i.e. when $k_\alpha = k_\beta + mK_x$, splitting comes out to be $\Delta = 2K_{\alpha\beta}$ in the $k_x$ component. It may be seen from Equation 4 that the azimuthal angle φ gives control on the coupling strength.

In Figure 2, we can observe anti-crossing feature between different modes and that the splitting related to the anti-crossing changes with φ. It has been shown earlier that the splitting in the energy and momentum could occur due to conservative and dissipative coupling, respectively [19]. As an example, we consider the case of interaction of (0,2) and (1,1) plasmon modes and how the interaction strength changes with $\varphi$. The theoretical values of the splitting have been calculated using equations 4 – 7 which are compared with the measured values in Table1 as a function of φ.

| Angle φ (degrees) | Experimental K-splitting $\Delta \pm 0.002$ ($\mu m^{-1}$) | Calculated K-splitting $\Delta$($\mu m^{-1}$) |
| --- | --- | --- |
| 15 | 0.043 | 0.0474 |
| 30 | 0.149 | 0.156 |
| 45 | 0.128 | 0.1338 |
| 60 | 0.048 | 0.0541 |

Table 1 Comparison of measured and calculated split gap for various azimuthal angles.

In summary, we have shown the anti-crossing of in-plane coupled plasmon modes in SPP dispersion. We also show that the strength of the coupling, manifested as the amount of split gap, can be controlled by the azimuthal angle (that is the measurement geometry).

**References:**

1. L. Novotny, Am. J. Phys. **78**, 1199 (2010).
2. R. M. Stevenson, V. N. Astratov, M. S. Skolnick, D. M. Whittaker, M. Emam-Ismail, A. I. Tartakovskii, P. G. Savvidis, J. J. Baumberg, and J. S. Roberts, Phys. Rev. Letts. **85**, 3680 (2000).
3. E. A. Stinaff, M. Schibner, A. S. Bracker, I. V. Ponomarev, V. L. Korenev, M. E. Ware, M. F. Doty, T. L. Reinecke, and D. Gammon, Nature **311**, 636 (2006).
4. K. F. MacDonald, Z. L. Samson, M. I. Stockman, N. I. Zheludev, Nature Photonics **3**, 55 (2009).
5. D. Sarid, R. T. Deck, and J. J. Fasano, J. opt. Soc. Am. **72**, 1345 (1982).
6. S. Dutta Gupta, G. V. Varada, and G. S. Agarwal, Phys. Rev. B **36**, 6331 (1987).
7. W. R. Holland and D. G. Hall, Phys. Rev. B **27**, 7765 (1983).
8. G. S. Agarwal and S. Dutta Gupta, Phys. Rev. B **32**, 3607 (1985).
9. J. Li, H. Lu, J. T. K. Wan and H. C. Ong, Appl. Phys. Letts. **94**, 033101 (2009).
10. A. Christ, T. Zentgraf, S. G. Tikhodeev, N. A. Gippius, O. J. F. Martin, J. Kuhl, and H. Giesen, Phys. Stat. Sol. (b) **243**, 2344-2348 (2006).
11. H. Gao, J. Henzie, M. H. Lee, and T. E. Odom, Proc. Natl. Acad. Sci. **105**, 20146-20151 (2008).
12. W. L. Barnes, W. A. Murray, J. Dintinger, E. Devaux, and T. W. Ebbesen, Phys. Rev. Letts. **92**, 107401 (2004).
13. A. Christ, S. G. Tikhodeev, N. A. Gippius, J. Kuhl, and H. Giessen, Phys. Rev. Letts. **91**, 183901 (2003).
14. H. Raether, *Surface plasmons on smooth and rough surfaces and on gratings*, (Springer-Verlag, Berlin 1986).
15. S. Dutta Gupta, J. Opt. Soc. Am. B **4**, 1893 (1987).
16. S. Kasture, P. Mandal, A. Singh, A. Ramsay, A. S. Vengurlekar, S. Dutta Gupta, V. Belotelov, A. V. Gopal, Appl. Phys. Letts. **101**, 091602 (2012).
17. H. Kogelnik, and C. V. Shank, J. Appl. Phys. **43**, 2327 (1972).

18. R. Dandliker, SPIE Proc. Fifth International Topical Meeting on Education and Training in Optics **3190** 279 (1997).
19. A. Kolomenskii, S. Peng, J. Hembd, A. Kolomenski, J. Noel, J. Strohber, W. Teizer, and H. Schuessler, Opt. Expr. **19**, 6587 (2011).
20. S. A. Maier, *Plasmonics: Fundamentals and Applications* (Springer, New York, USA 2007).
21. Jia-ming Liu, *Photonic Devices* (Cambridge University Press, UK 2005).